\documentclass[preprint,showpacs,preprintnumbers,amsmath,amssymb]{revtex4}

\usepackage{graphicx}
\usepackage{dcolumn}
\usepackage{bm}

\begin{document}

\preprint{APS/123-QED}

\title{Ion condensation on charged patterned surfaces}%

\author{Yuri S. Velichko}%
\affiliation{Department of Materials Science and Engineering,
Northwestern University, Evanston, IL 60208, USA.}%

\author{Francisco J. Solis}%
\affiliation{Arizona State University West, Integrated Natural Sciences,
Phoenix, AZ 85069, USA.}%

\author{Sharon M. Loverde}%
\affiliation{Department of Materials Science and Engineering,
Northwestern University, Evanston, IL 60208, USA.}%

\author{Monica Olvera de la Cruz}%
\email{m-olvera@northwestern.edu}%
\affiliation{Department of Materials Science and Engineering,
Northwestern University, Evanston, IL 60208, USA.}%
\date{\today}

\begin{abstract}
We study ion condensation onto a patterned surface of alternating charges. The
competition between self-energy and ion-surface interactions leads to the
formation of ionic crystalline structures at low temperatures. We consider
different arrangements of underlying ionic crystals, including single ion
adsorption, as well as the formation of dipoles at the interface between
charged domains. Molecular dynamic simulation illustrates existence of single
and mixed phases. Our results contribute to understanding pattern recognition,
and molecular separation and synthesis near patterned surfaces.

\end{abstract}



\maketitle

The interaction of molecules with surfaces or interfaces is a fundamental
problem in modern science. Advances in molecular design and maturity of
nanoscale microscopy paved the way in a number of novel technological
achievements, such as the ability to produce nano-structured
surfaces~\cite{AndelmanScience1995,GrovesPRL2005,MirkinScience1999}, and
complex self-assembled
aggregates~\cite{KalerJCP1992,ZembNature2001,StuppScience2001,Olvera}.
Periodic and spatially organized surfaces can be used as templates for
fabricating nanostructures, magnetic storage media,
nanowires~\cite{JJdePabloNature2003}, or substrates for control of cell
adhesion and growth~\cite{GrovesLangmuir2001,Sackmann2005}. From a scientific
point of view, charged surfaces and their interactions with ionic environments
are particularly interesting due to the long-range nature of electrostatic
interactions and its relative independence from molecular structure. The break
down of the mean field Poison-Boltzmann approximation to describe adsorption
of ions to charged surfaces~\cite{MarceljaCPL1984} motivated many studies.
Simple models such as a single component plasma near a neutralizing background
revealed two-dimensional crystallization of adsorbed ions onto homogeneously
charged surfaces~\cite{RouzinaJPhysChem1996}. That is, the electrostatic
repulsion between condensed ions results in lateral correlations.

\begin{figure}
\includegraphics[width=8.5cm]{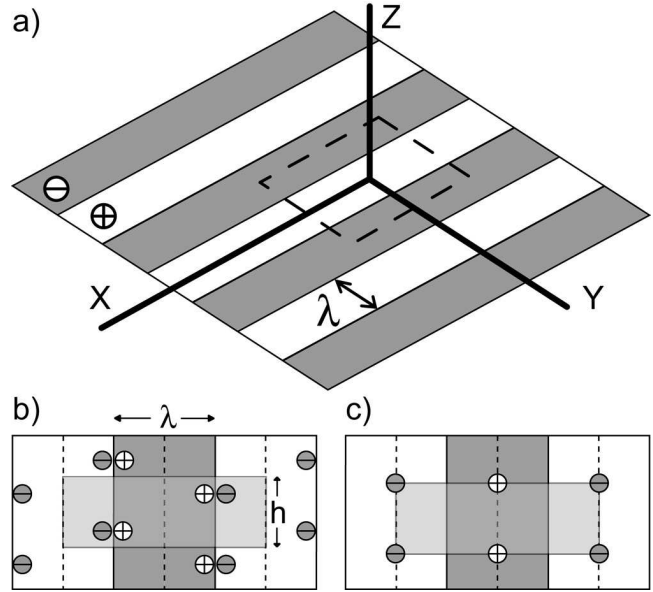}
\caption{(a)~The schematic representation of the system. The basic Wigner
lattices formed by condensed (b)~ions forming dipoles and (c)~single ions.}
\label{Model}
\end{figure}

In this letter we study ion condensation on a planar surface with periodic
domains of alternating charge. In particular, we consider striped patterns of
width $\lambda$ and constant surface charge number density $\sigma$
(Fig.~\ref{Model}(a)). The half-space above the plane is filled by a solution
of positive and negative ions with valency $Z$, radius $a$ and bulk number
density $\rho^{o}=\rho^{o}_{+}=\rho^{o}_{-}$. We consider a symmetric system
with ions of equal valency. At low temperatures, the homogeneously charged
regions of the surface attract oppositely charged ions. The long-range
electrostatic interaction induces periodic arrangement of condensed ions
within the plane in analogy with Wigner crystals~\cite{Madelung}. The
competition between ion-ion and ion-surface electrostatic interactions,
together with the geometrical restrictions set by the surface patterns
determine the structure of the ionic crystal. When the ion-ion interaction
dominates, we find the formation of dipolar pairs along the interfaces between
the surface patterns (Fig.~\ref{Model}(b)). On the other hand, when the
ion-surface interaction dominates, the ions localize where the electrostatic
potential is strongest, at the center of the stripes (Fig.~\ref{Model}(c)). It
is the goal of this letter to formulate an approach that captures the
essential physics of the phenomenon and gives a reasonable description of ion
condensation near patterned surfaces.

We model the adsorption and the interaction of the ions between themselves and
with the surface as follows. The free energy of the absorbed ions is purely
electrostatic and includes contributions from interactions between the ions
and surface as well as between ions. The free energy of the ions in the bulk
is dominated by their translational entropy. We note that the bulk ions are
screened from the surface by the condensed ions within the Gouy-Chapman
length~\cite{RouzinaJPhysChem1996}, $l_{o}=1/2\pi\sigma Z\ell$. This allows us
to consider the layer of adsorbed ions separately form the bulk. Equilibrium
between the two regions is achieved when the chemical potentials of condensed
ions, $\mu_c$, and bulk ions, $\mu$, are equal $\mu_c=\mu$. Equivalently, this
condition can be stated as the requirement that the excess energy of the
condensed ions $F$, per unit area $A$, be a minimum with respect to the
condensed ion density and arrangement. We write this effective free energy as:
\begin{eqnarray}\label{FreeEnergy}
\frac{1}{A}\frac{F(\rho,\mu)}{k_BT}= \rho\left(\frac{\ell
Z^2}{2}\frac{M}{n}+\upsilon+\varepsilon-\mu\right),
\end{eqnarray}
where $\ell=e^2/4\pi\epsilon_{\rm o}k_BT$ is Bjerrum length, $\rho$ is a
surface number density of condensed ions, $M$ is the Madelung constant of a
crystal of condensed charged units, $n$ is a number of ions per unit,
$\upsilon$ is the energy of ion-surface interaction, and $\varepsilon$ is the
self-energy of the condensed units. The condensed unit can be single ion
($n=1$) with zero self energy, $\varepsilon=0$, dipole composed of two
opposite charges ($n=2$), with $\varepsilon=-Z^2\ell/a$, or more complex
structure like a chain or a cluster of ions.

The minimum of the free energy is determined by evaluating and
comparing the values of expression Eq.~\ref{FreeEnergy} for
different ion densities and lattice symmetries. The key step in
these evaluations is the determination of the Madelung constant for
the basic arrangements of ions (Fig.~\ref{Model}(a,b)). This
constant is the effective potential experienced by a charged unit
due to the presence of a lattice of similar units. For the system of
dipoles (Fig.~\ref{Model}(b)), we have
\begin{equation}
M_{d}=-\acute{\sum\limits_{ij}}\,\vec{p}\;\vec{E}=
\acute{\sum\limits_{ij}}(-1)^{i+1}
\left[\frac{(3\vec{p}\;\vec{r}_{ij})^2}{r^5_{ij}}-\frac{p^2}{r^3_{ij}}\right]
\end{equation}
where $i$ and $j$ are lattice indices, $\vec{p}$ is a dipole momentum of
magnitude $p=1/a$, $\vec{E}_{ij}$ is the electric field at the position
$\vec{r}_{ij}=\left( h(j+i/2),\lambda i\right)$. The sum over integers $i$ and
$j$ runs from $-\infty$ to $+\infty$ and the prime over summation indicates
the avoidance of singularity at the vector $(i,j)=(0,0)$. Due to the constant
width of the stripe, the separation distance between condensed ions along the
stripe can be written as $h=n_{+}/\rho_{+}\lambda$, where $n_{+}$ is the
number of positively charged ions in the Wigner crystal cell. For the case of
ion absorption to the center of the stripe (Fig.~\ref{Model}(c)), we obtain:
\begin{equation}
M_{c}=-\acute{\sum\limits_{ij}}\,\psi_{ij}=
\acute{\sum\limits_{ij}}(-1)^{i+1}\frac{1}{r_{ij}},
\end{equation}
where $\psi_{ij}$ is the potential at position $\vec{r}_{ij}=\left( jh,\lambda
i\right)$. These sums can be calculated using Ewald summations~\cite{Ewald},
following, for example, the methods developed by Crandall~\cite{Crandall}. For
the dipolar case, which does not pose convergence problems, faster techniques
are available~\cite{Lekner,Sperb}. Besides these two basic arrangements, we
have also calculated Madelung constants for lattices with a relative
translation along the stripe for adjacent lines of ions. We find that both
lattices shown in Fig.~\ref{Model}(b,c) have the lowest energies.

The electrostatic potential near the striped surface (Fig.\ref{Model}(a)) is
given by
\begin{eqnarray}
\Psi(\Delta y,z)=e\sigma\!\!\int\limits_{-\infty}^{\infty}\!\!dx\!\!
\int\limits_{-\lambda/2}^{\lambda/2}\!\!dy\!\!
\sum\limits_{i=-\infty}^{\infty}\psi_i(x,y,\Delta y,z)
\end{eqnarray}
where $e$ is an electron charge (negative), $\sigma$ is a surface number
charge density, $\lambda$ is the width of a stripe and
\begin{eqnarray}
\psi_i(x,y,\Delta y,z)=\frac{(-1)^{i}}{\sqrt{x^2+(y-\Delta
y+i\lambda)^2+z^2}},
\end{eqnarray}
where $\Delta y\in[-\lambda/2,\lambda/2]$ determines the shift from the center
of the stripe along the $y-$axis. Using the identity~\cite{Sperb} the
electrostatic potential can be calculated exactly:
\begin{eqnarray}\label{surf_potential}
\Psi(\Delta y,z)=\frac{8e\sigma\lambda}{\pi}\times\nonumber\\
\sum\limits_{l=1}^{\infty}\frac{\sin^3\left(\pi l/2\right)}{l^{2}}
\sin\left(\frac{\pi l(\lambda-2\Delta y)}{2\lambda}\right) e^{-\frac{\pi
lz}{\lambda}}.
\end{eqnarray}
The magnitude of the potential decays exponentially and has sinusoidal profile
with extremum in the center of the stripes. Thus, the energy of the single
charge in the center of the stripe (Fig.~\ref{Model}(c)) is $\upsilon_c=-4\pi
Ze^2\sigma a\lambda/3$ and the energy of the dipole at the interface
(Fig.~\ref{Model}(b)) is $\upsilon_d=-2\pi Ze^2\sigma a^2.$

In order to investigate the phase behavior of the system first we determine
the conditions for ion adsorption into each of the two states. The adsorption
of ions into a regular structure starts when the bulk chemical potential is
lower than the critical value implicitly determined from the equation:
\begin{equation}\label{mu_crtc}
\frac{\partial}{\partial\rho}
\left.\left(\frac{1}{A}\frac{F(\rho,\mu)}{k_BT}\right)\right|_{\rho=0}=0,
\end{equation}
when we consider the free energy as a function of the chemical potential. It
is equivalent to appearance of the first non-trivial extremum of the free
energy. This conditions (Eq.\ref{mu_crtc}) can be solved exactly for two types
of condensed ionic lattices. In the limit of small surface charge density, the
adsorption starts from the formation of the dipoles along the boundary of the
surface patterns (Fig.~\ref{Model}(b)), due to the weak interaction with the
surface. The critical chemical potential for the system of condensed dipoles
reads
\begin{equation}\label{mu_crtc_A}
\mu_{cr}^A=-\ell Z\left(\frac{2\pi\sigma a^2+Z}{2a}\right).
\end{equation}
It is important to note that it is $\lambda$ independent. With further
increase in the surface charge density, the energy of the electrostatic
attraction to the surface overcomes the energy of the dipole formation and the
ionic crystal of condensed ions changes its structure. The new ionic crystal
(Fig.~\ref{Model}(c)) has a simple rectangular lattice and is formed from the
single ions located in the minima of the surface electrostatic potential. The
critical chemical potential in that case is
\begin{eqnarray}\label{mu_crtc_B}
\mu_{cr}^B=-\ell Z\left(\frac{4\pi\sigma\lambda^2+3Z\log[2]}{3\lambda}\right)
\end{eqnarray}
The transition from one structure to another occurs at critical surface charge
density
\begin{equation}\label{sigma_cr}
\frac{\sigma_{cr}}{Z}=\frac{3}{2\pi a\lambda}
\left(\frac{\lambda-2a\log[2]}{4\lambda-3a}\right),
\end{equation}
when the energy of both systems are equal, or equivalently, both have the same
chemical potential. The critical surface charge density determines the
beginning $(\sigma_{cr},\mu_{cr})$ of the coexistence line $\mu_{tr}$, which
separates regimes for condensed ions forming ionic crystals of dipoles
(Fig.~\ref{Model}(b)) from the one of the ionic crystals of single charge
(Fig.~\ref{Model}(c)).

Once one of the condensation conditions $\mu<\mu_{cr}$ is satisfied,
the associated ions form structures with well defined densities
$\rho$. In these conditions, the density $\rho$ of the condensed
ions can be calculated from
\begin{equation}
\frac{\partial}{\partial\rho}\left(\frac{1}{A}\frac{F(\rho,\mu)}{k_BT}\right)=0,
\end{equation}
for the corresponding lattice symmetry. The preferred symmetry of the ionic
crystal is determined by direct comparison of the two energy minima achieved
by each of the structures considered. We note that at the transition between
structures, when both energies are equal, the associated density is different,
and the transition between phases is first order.

The stability of phases A and B is determined as follows. Phase A is composed
of dipolar units, and in general, the addition of one more dipole does not
reduce the energy. However, the introduction of an isolated ion at the center
of an oppositely charged stripe might indeed reduce the energy, signaling the
fact that phase A is no longer the minimum-energy conformation. Similarly, we
consider the addition to a background of stripe centered ions, phase B, of a
single dipole at the interface between two oppositely charged stripes. When
these perturbations reduce the energy of the phase with least energy (between
A and B), the actual minimum of that state corresponds to a more complex
structure. It is likely that the actual minimum structure is a combination of
both striped-centered ions and dipoles, denoted here as a {\it mixed} phase.
For example, for the dipole phase A, the instability criteria reads:
\begin{equation}\label{delta_F}
\frac{\delta F}{k_BT}=\frac{\ell Z^2}{2}M_{cA}+\upsilon_{c}-\mu<0,
\end{equation}
where $\upsilon_{c}$ is the energy of the ion-surface interaction, and
\begin{equation}
M_{cA}=2\pi a\rho_{A}+ 16\pi
a\lambda^2\rho_{A}^2\sum\limits_{l=1}^{\infty}lK_1\!\!\left[\pi
l\lambda^2\rho_{A}\right]\cos\left(\pi l\right)
\end{equation}
is the energy of a single ion at the center of the stripe in the
presence of the dipolar array (Fig.~\ref{Model}(b)), and $K_1$ is a
modified Bessel function of first order. The spinodal boundary (when
the system is first unstable), appears at values of the chemical
potential $\mu_{sp}^A$ for which $\delta F=0$. The spinodal boundary
for the stripe-centered case is labelled $\mu_{sp}^B$.

\begin{figure}
\includegraphics[width=8.5cm]{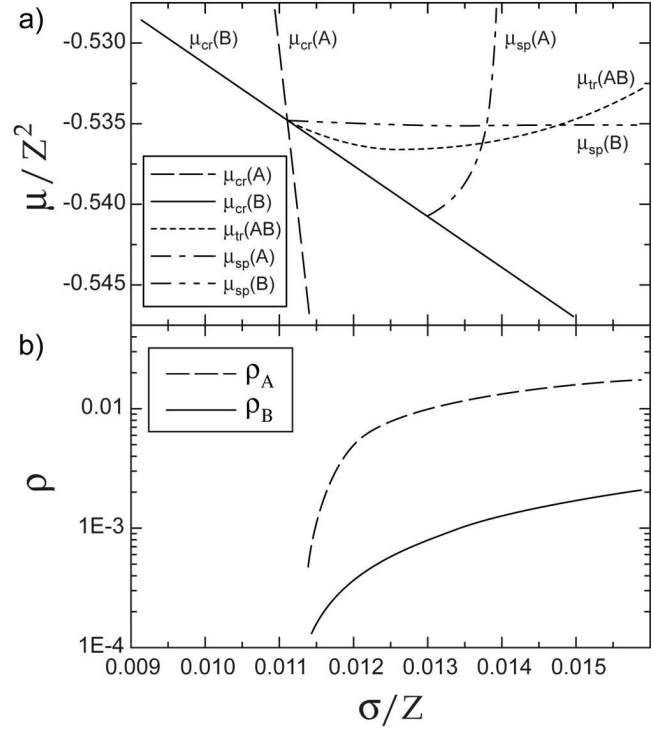}
\caption{(a) The dependence of the critical chemical potential ($\mu_{cr}$),
coexistence ($\mu_{tr}$) and spinodal ($\mu_{sp}$) lines vs. surface charge
density $\sigma/Z$. (b) The dependence of the density vs. the surface charge
density for both Wigner lattices formed by dipoles ($\rho_A$) and single ions
($\rho_B$) along the coexistence line.} \label{Results}
\end{figure}

\begin{figure}
\includegraphics[width=8.5cm]{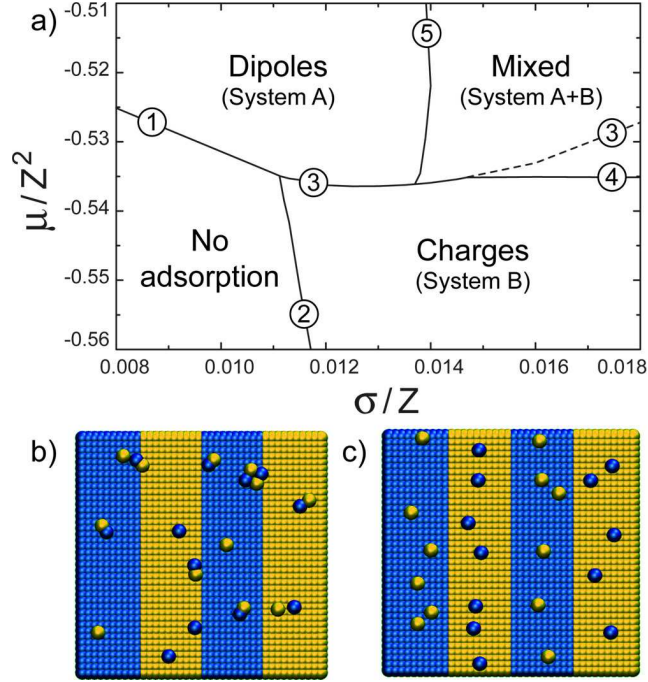}
\caption{(a) The schematic phase diagram implied by results of calculation,
where lines $(1)$ and $(2)$ correspond to $\mu_{cr}^A$ and $\mu_{cr}^B$, line
$(3)$ is a coexistence line $\mu_{tr}$, line $(4)$ and $(5)$ are spinodal
lines $\mu_{sp}^B$ and $\mu_{sp}^A$ correspondingly. (b, c) Snapshots from
molecular simulation of mixed and single ion phases at $T=0.5k_BT$.}
\label{PhaseDiag}
\end{figure}

We illustrate these results for the cases $\lambda/a=10$ and
$Z\ell=1$. Figure~\ref{Results}(a) shows the dependence of the
critical chemical potential vs. surface charge density $\sigma$ for
systems of both symmetries (Eqs.~\ref{mu_crtc_A}
and~\ref{mu_crtc_B}). The condition of equal chemical potential
between phases A and B determines the coexistence boundary line
$(\mu_{tr})$. Figure~\ref{Results}(a) also shows the location of the
spinodal curves for the dipole $\mu_{sp}^{A}$, and the centered
$\mu_{sp}^{B}$ states. Along the coexistence line between phases A
and B, the system undergoes a first order phase transition
associated with the jump in the density of condensed ions.
Figure~\ref{Results}(b) shows the dependence of the density for both
systems along the coexistence line. Figure~\ref{PhaseDiag}(a)
summarizes results in a schematic phase diagram in terms of the
surface charge density $\sigma/Z$ and the chemical potential
$\mu/Z^2$ normalized by the valency $Z$ of the ions.

Figure~\ref{Results}(a) shows two basic features of the boundaries of
stability for the dipolar and centered structures. Note that the spinodal line
for the stripe centered state, system B, lies approximately at a constant
chemical potential. At lower values of the potential, it is simply more
convenient to add single ions at the center of the stripe. Reciprocally, the
limit of stability of the dipolar state, system A, appears at near constant
value of the surface charge. This surface charge density determines the
strength of the electric field at the interface between stripes, and sets the
energy gain for a deposited dipole. For strong enough values of the charge
density, the energy of the stripe centered state can be lowered by adding
dipoles at the boundaries. Large values of both, chemical potential and charge
density, lead therefore to the complex mixed state in the upper right corner
of our phase diagram.

We explore finite temperature effects in the phase diagram with molecular
dynamics simulations. Condensed ions form dipoles along the interfaces and
single ions lie in the center (Fig.\ref{PhaseDiag}(b,c)). The strong
correlation between ions, is preserved even at relatively high temperature
$T=0.5k_BT$; thermal fluctuations do not destroy the long range order. In our
simulations we have found, in addition to dipole and single charge phases,
chain clusters and paired dipoles suggesting rather complex phase behavior
previously found mostly in the bulk~\cite{FisherPhysicaA1996,dePabloPRL2002}.

Our exploration of the low temperature limit of ion condensation onto a
structured surface shows that there are many different controllable behaviors
that a fixed surface pattern can generate. It seems plausible to use the
properties of condensed ions to create complex, tunable structures at
interfaces that exhibit a variety of potential applications. For example,
electrostatic charge on membrane surface can be used to attract free vesicles
or membrane-coated charged microbeads~\cite{GrovesLangmuir2003} to employ
intermembrane adhesion and lipids exchange with substrate-supported patterned
membrane. Another example, a new more effective and selective method, compared
to previous DNA fractionation techniques, was demonstrated using streaming
dielectrophoresis of DNA on the Si surface with micro-scaled Au
strips~\cite{RafailovichPRL2007}. Control of the abundance of the charged
species can lead to dramatic transitions, that can be used as catalytic steps
of complex chemical reactions, as steps in the formation of secondary
self-assembled structures superimposed on a simpler pattern. Though our
analysis is focused only on two basic behaviors, stripe-centered, and dipolar
arrangements, we show the existence of more complicated, mixed, phases. We
expect, in fact, a cascade of transitions between all these complex phases.

This work is supported by NSF grant numbers DMR-0414446 and DMR-0076097. The
authors thank G. Vernizzi for helpful and stimulating discussions.

\newpage

\end{document}